\documentclass{article}
\usepackage[utf8]{inputenc}
\usepackage{amssymb}
\usepackage[english]{babel}
\usepackage{amsmath}
\usepackage{graphicx}
\usepackage{bm}
\usepackage{wrapfig}
\usepackage{sidecap}
\usepackage[format=plain, labelfont=it, textfont=it]{caption}
\usepackage[affil-it]{authblk}
\usepackage{cite}

\begin{document}
\title{Deep Neural Networks for Computational Optical Form Measurements}
\author{Lara Hoffmann%
  \thanks{Electronic address: \texttt{lara.hoffmann@ptb.de}; Corresponding author}$\ $ and Clemens Elster}
\affil{Physikalisch-Technische Bundesanstalt, Braunschweig and Berlin, Germany}

\date{Dated: \today}

\maketitle

\begin{abstract}
Deep neural networks have been successfully applied in many different fields like computational imaging, medical healthcare, signal processing, or autonomous driving. In a proof-of-principle study, we demonstrate that computational optical form measurement can also benefit from deep learning. A data-driven machine learning approach is explored to solve an inverse problem in the accurate measurement of optical surfaces. The approach is developed and tested using virtual measurements with known ground truth.
\end{abstract}


\section{Introduction}  
Deep neural networks and machine learning in general are experiencing an ever greater impact on science and industry. Their application has proven beneficial in many different domains, including autonomous driving \cite{grigorescu}, anomaly detection in quality management \cite{staar}, computational imaging \cite{barbastathis}, signal processing \cite{mousavi}, analysis of raw sensor data \cite{moraru}, or medical health care \cite{esteva}, \cite{kretz}. Machine learning methods have also been successfully employed in optics. Examples comprise the compensation of lens distortions \cite{chung}, or correcting abberated wave fronts in adaptive optics \cite{vdovin}.

Machine learning has been used for misalignment corrections \cite{baermiss}, \cite{zhang}, aberration detection \cite{yan}, or phase predicitons \cite{rivenson2}. But to the best of our knowledge, deep learning has not yet been applied for the accurate computational measurement of optical aspheres and freeform surfaces predicting the surface under test from its optical path length differences. The precise reconstruction of aspheres and freeform surfaces is currently limited by the accuracy of optical form measurements with an uncertainty range of approximately $50$ nm \cite{schachtschneider}. The aim of this paper is to demonstrate through a proof-of-principle study that this field in optics can also significantly benefit from machine learning techniques. 

\begin{figure}[h]
  \centering
  \begin{minipage}[t]{0.52\textwidth}
    \centering
    \includegraphics[scale = 0.5]{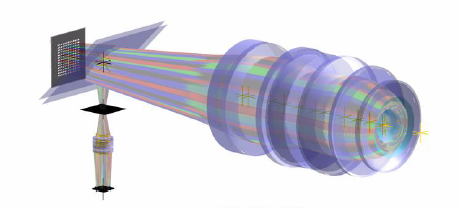}
    \caption{Schematic of the tilted-wave interferometer (reference arm not shown). The $2$D point source array is on the left, the specimen is on the right and the CCD is at the bottom.}\label{twischeme}
  \end{minipage}
  \hfill
  \begin{minipage}[t]{0.44\textwidth}
    \centering
    \includegraphics[scale = 0.35]{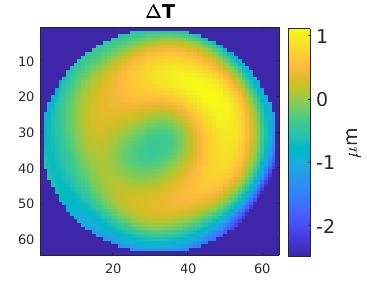}
    \caption{An example of a difference topography $\Delta T$ on a $64\times 64$ pixel grid.}\label{difftopo}
    
  \end{minipage}
\end{figure}

Our investigations were conducted using the SimOptDevice \cite{simopt} simulation toolbox. The toolbox provides realistic, virtual experiments with known ground truth. We concentrated on the tilted-wave interferometer (TWI) \cite{baer} for the experimental realization. It is a promising technique for the accurate computational measurement of optical aspheres and freeform surfaces using contact-free interferometric measurements. The TWI combines a special measurement setup with model-based evaluation procedures. Four CCD images with several interferograms are generated by using multiple light sources to illuminate the surface under test. A simplified scheme is shown in Figure \ref{twischeme}. The test topography is then reconstructed by solving a numerically expensive nonlinear inverse problem by comparing the measured optical path length differences to simulated ones using a computer model and the known design of the surface under test. In this study, PTB's realization of the TWI evaluation procedure is considered \cite{ptbtwi}.

While the great success of deep networks is based on their ability to learn complex relations from data without knowing the underlying physical laws, including existing physical knowledge into the models can further improve results (cf. \cite{bezenac}, \cite{karpatne} or \cite{raissi}). In our study, we also follow such an approach by developing a hybrid method which combines physical knowledge with data-driven deep neural networks. The employed scientific knowledge is twofold; training data is generated by physical simulations and a conventional calibration method is used to generalize the trained network to non-perfect systems.

This paper is organized as follows. Section $2$ briefly introduces neural networks and presents the proposed deep learning framework. The means of generating the training data and the details of training the network are explained, combining this approach with a conventional calibration method for better generalization. The results obtained for independent test data are then presented and discussed in Section $3$. Finally, some conclusions are drawn from our findings and possible future research is suggested.

\section{Hybrid method}
This section provides an overview of deep neural networks and introduces the hybrid method that was developed which combines the TWI procedure with a data-driven deep learning approach. Without loss of generality, each specimen can be assumed to have a known design topography. The overall goal of form measurement is to determine the deviation $\Delta T$ of a specimen $T_s$ to the given design topography $T_d$, i.e. : $ T_s = T_d + \Delta T$. To this end, the TWI provides measurements of the optical path length differences $L_s$ of the specimen under test. Simultaneously, a computer model \cite{simopt} computes the optical path length differences $L$ of a given topography $T$. The inverse problem is to reconstruct the specimen topography $T_s = T_d + \Delta T$ from its measured optical path length differences $L_s$.

The conventional evaluation procedure of the TWI is numerically expensive and relies on linearization. A general advantage of neural networks is the ability to produce instant results once they are trained. Furthermore, it is interesting to explore whether deep learning could also improve the quality of the inverse reconstruction as a nonlinear approach. We aim to address the inverse problem described above using deep networks, i.e. by reconstructing a difference topography $\Delta T$ from given differences of optical path length differences $\Delta L = L_s - L_d$.

\subsection{Data generation}

\begin{figure}[h]
    \centering
    \includegraphics[scale = 0.22]{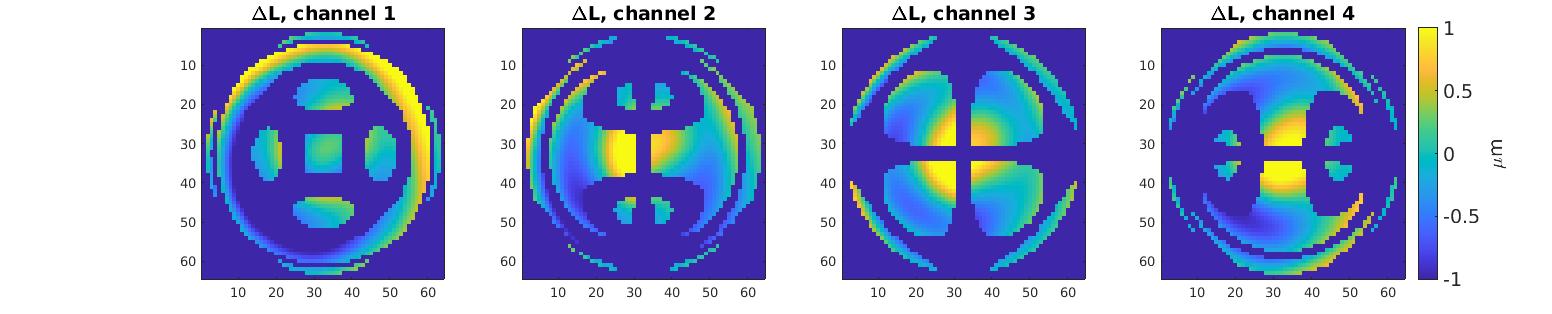}
    \caption{An example of the calculated differences of optical path length differences for one specimen $\Delta L$ with an asphere as the underlying design topography. The four different images originated from the disjoint sets of wave fronts that resulted from the four different mask settings at the $2$D point source array.}\label{opdd}
\end{figure}

When solving an inverse problem with neural networks, it is a common practice to generate data through physical simulations \cite{lucas}, \cite{mccann}. Here, various difference topographies $\Delta T$ are generated through randomly chosen weighted Zernike polynomials. They are then added to a specific design topography at a fixed measurement position to create different virtual specimens. The sequence of Zernike polynomials yields an orthogonal basis of the unit disc and is a popular tool in optics to model wave fronts \cite{wang}. Following the forward pass, the computer model is used to compute optical path length differences of the design topography and the modeled specimens. Note that the data is generated assuming perfect system conditions, i.e. the computer model is undisturbed.  An example can be seen in Figures \ref{difftopo} and \ref{opdd}.

Data were generated for two different design topographies with about $22.000$ data points for each design. It should be noted, that $10$\% of the data was used exclusively for testing and was not included in the network training.

\begin{figure}[h]
  \centering
  \begin{minipage}[t]{0.46\textwidth}
    \centering
    \includegraphics[scale = 0.25]{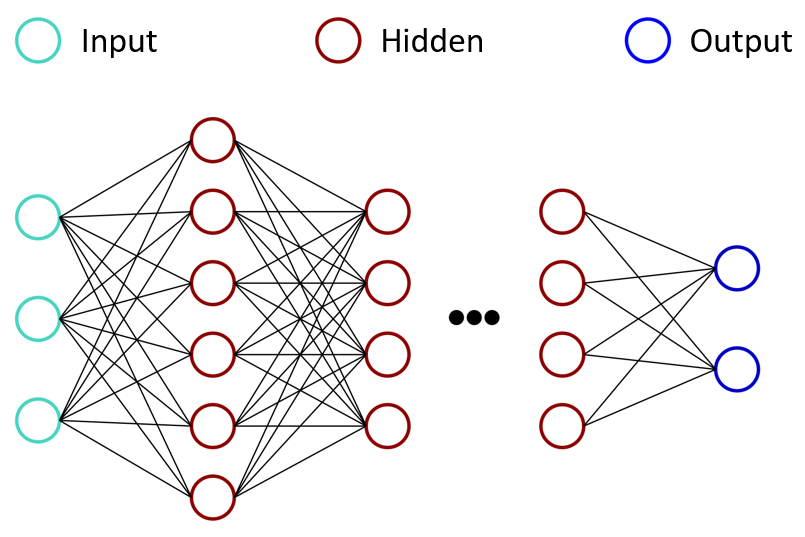}
    \caption{An example of a deep neural network architecture.}\label{dnn}
  \end{minipage}
\hfill
  \begin{minipage}[t]{0.5\textwidth}
    \centering
    \includegraphics[scale = 0.35]{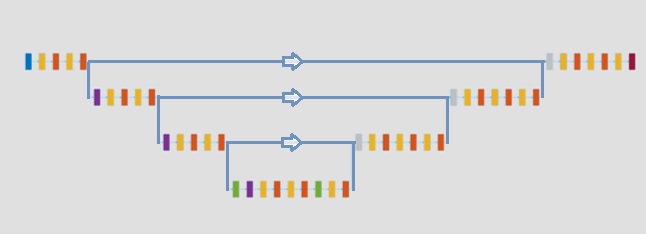}
    \caption{An example of a U-Net structure.}\label{unet}
  \end{minipage}
\end{figure}

\subsection{Deep neural network architecture}
A simple, fully connected neural network with a single hidden layer is represented by a nonlinear function $f_\Phi : \mathbb{R}\rightarrow\mathbb{R}$ with parameters $\Phi = \lbrace \omega_i,b_i\in\mathbb{R}|i=1,\ldots,n\rbrace$, where $n$ is the number of neurons in the hidden layer. The univariate output of the network is modeled as $f_\Phi (x) = \sum_{i=1}^n\sigma (\omega_ix+b_i),\ x\in\mathbb{R},$ where $\sigma$ is a nonlinear activation function. In general, input and output are higher dimensional, and the architecture can become arbitrarily deep by adding more layers. Neural networks with high complexity are called "deep neural networks". An example of this type of architecture is shown in Figure \ref{dnn}. There, two outputs are predicted based on three given inputs after processing the information through several hidden layers. Also, different types of layers - convolutional layers \cite{cun}, for example- can be used instead of fully connected ones. The network parameters can be optimized via backpropagation on the given training data by minimizing a chosen loss function between the predicted and known output.

At this point, it is important to consider the network as being an image-to-image regression function $f_\Phi$ which maps the differences of optical path length differences $\Delta L$ (see Fig. \ref{opdd}) onto a difference topography $\Delta T$ (see Fig. \ref{difftopo}), i.e.: $f_\Phi:\ \mathbb{R}^{M\times M\times K}\rightarrow\mathbb{R}^{M\times M},\ \Delta L\mapsto\Delta T,$ where $\Phi$ are the network parameters to be trained, $M\times M$ is the dimension of the images, and $K$ is the number of channels in the input. Note that the image dimension of the input equals the image dimension of the output here. This is not mandatory, but it suits the network architecture described below very well. While the CCD gives a resolution of $2048\times 2048$ pixels, we chose $M=64$. For the asphere and the multi-spherical freeform artefact \cite{multisphere} as seen in Figure \ref{opdd}, $K=4$ and $K=1$ were used, respectively. This is because the multi-spherical freeform artefact has a big patch in the first channel which almost covers the entire CCD for the selected measurement position. Furthermore, even though some pixels are missing, the first channel sufficed for the purpose of this deep learning proof-of-principle study.

We chose a U-Net as the network architecture. U-Nets have been successfully applied in various image-to-image deep learning applications \cite{isil}. An example of a structure is shown in Figure \ref{unet}. The input passes through several convolution and rectified linear unit layers on the left side before being reduced in dimension in every vertical connection. After reaching its bottleneck at the bottom, the original data dimension is restored step by step through transposed convolution layers on the right side. During each dimensional increase step, a depth concatenation layer is added which links the data of the current layer to the data of the former layer with same dimension. These skip connections are depicted as horizontal lines in Figure \ref{unet}.

Here, the chosen U-Net architecture consists of a total of $69$ layers. The training set was used to normalize all input and output data prior to feeding them into the network. The U-Net was trained using an Adam optimizer \cite{adam} and the mean squared error as the loss function. About two hours of training were carried out for the multi-spherical freeform artefact with an initial learning rate of $0.0005$, a drop factor of $0.75$ every five periods, and a mini batch size of $64$. In addition, a dual norm regularization of the network parameters with a regularization parameter of $0.004$ was employed to stabilize the training. For the asphere, training was carried out for $15$ epochs with a mini batch size of eight samples, an initial learning rate of $0.0005$ which decreased every three epochs by a learning rate drop factor of $0.5$, and a regularization parameter of $0.0005$.

\subsection{Generalization to non-perfect systems}
In real world applications there no perfect systems exist. Thus, the computer model needs to be adapted phenomenologically. In the conventional calibration procedure, the beam path is calibrated through the computer model by using known, well fabricated spherical topographies at different measurement positions to compare the optical path length differences measured by the TWI and its computer model \cite{fortmeier}. 

\begin{figure}[h]
  \centering
\includegraphics[scale = 0.6]{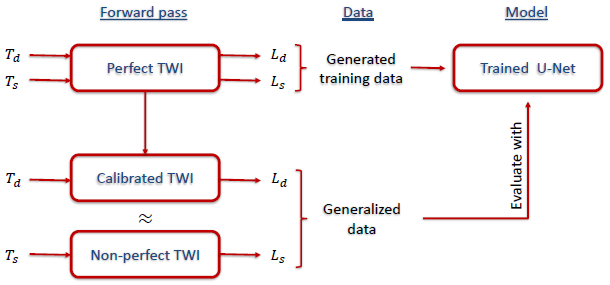}
 \caption{Workflow of the developed hybrid method. }\label{hybrid}
\end{figure}

It is not feasible to generate an entire new data base and train a new network after each system calibration when using deep networks. Nonetheless, the ultimate goal is to apply the trained deep network to real-world data. To this aim, we propose a hybrid method that trains the selected U-Net on data generated under perfect system conditions but also generalizes well to non-perfect systems by evaluating data derived through the conventional calibration method. A workflow chart of the hybrid method is shown in Figure \ref{hybrid}.

\section{Results}
The following results are all based on simulated data. As mentioned above, two different design topographies are considered as mentioned above: an asphere and a multi-spherical freeform artefact. First, the results of data acquired from a perfect system environment are presented. The networks which were trained for the design topography of an asphere and a multi-spherical freeform artefact are addressed, respectively. Next, additional strategies which could improve the models are discussed as well. Finally, the application of the hybrid method which was developed is presented in a non-perfect system environment.

The topographies have a circle as the base area. Since the required input and output of the network are images, the area outside of the circle shape is defined with zeros which the network learns to predict. Nonetheless, only the difference topography pixels inside the circle shape are considered in the presented results. 

\subsection{Perfect system}
\begin{figure}[t]
    \centering
    \includegraphics[scale = 0.4]{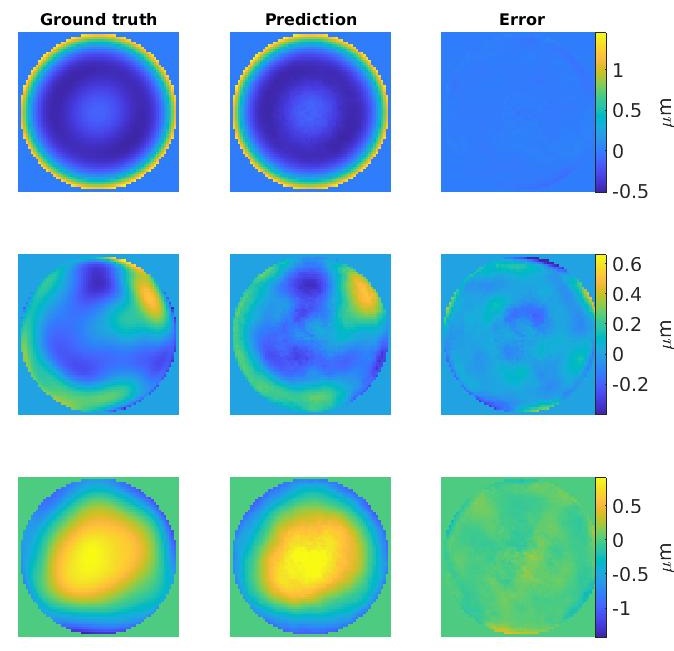}
    \caption{Three examples of predicted test difference topographies. The errors in the third column show the differences between the ground truth and the prediction. The root mean squared errors from top to bottom are $16$ nm, $84$ nm, and $74$ nm.}\label{exval}
\end{figure}

About $2200$ samples were used for testing. They were not included in the training. First, the multi-spherical freeform artefact was considered as the design topography. Three randomly chosen prediction examples are shown in Figure \ref{exval}. The root mean squared error of the U-Net predictions on the test set is $33$ nm. For comparison, the difference topographies in the test set have a total root mean squared deviation of $559$ nm. The median of the absolute errors of the U-Net is about $18$ nm, while the median of total absolute deviations in the test set is $428$ nm.

For the asphere as the design topography the root mean squared error is $102$ nm, while the test set has a root mean squared deviation of $589$ nm. The median of the absolute errors of the U-Net is $52$ nm and the median absolute deviation of the test set is $451$ nm for comparison. One possible explanation for the discrepancy in the accuracy of the predictions between the network for the asphere and multi-spherical freeform artefact as the design topographies is the following. The input of the respective U-Nets and their resulting architecture vary widely. As mentioned above, the network concerning the asphere has four input channels. These can be seen in Figure \ref{opdd}. In each channel, various different areas are illuminated at the CCD, resulting in a distribution of information into different and smaller patches. The multi-spherical freeform artefact on the other side, illuminates one big circle shaped patch in the first channel for the selected measurement position. This channel, which contains most of the important information in a single patch, forms the only input to the corresponding network.

\begin{figure}[h]
    \centering
     \includegraphics[scale = 0.35]{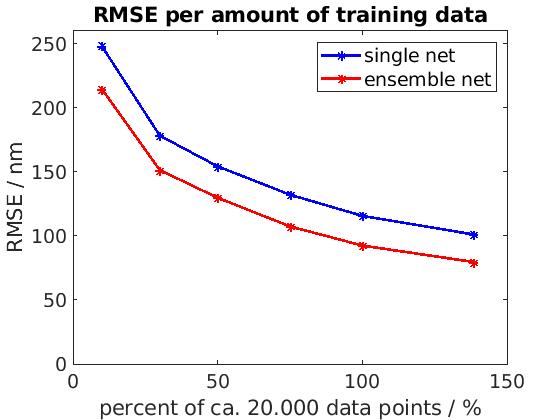}
    \caption{Root mean squared error of a single network (blue) and a network ensemble (red), depending on the amount of data used for training. The test set is the same for all evaluations and was not included in training.}\label{amounttestdata}
\end{figure}

However, the results for the asphere can be improved further. One way to do so is to increase the amount of training data (cf. Fig. \ref{amounttestdata}). As the input has four channels for the asphere, it seems natural that more data is needed for training than for the multi-spherical freeform artefact. A second approach is to use a network ensemble \cite{zhou} rather than a single trained network. To this end, $15$ U-Nets were trained from scratch and the ensemble output was taken as the mean of the ensemble predictions. The results are shown in Figure \ref{amounttestdata}. In this way, the accuracy was already improved to a root mean squared error of $80$ nm using an ensemble of $15$ U-Nets, each trained from scratch on almost $28.000$ data points. It should be noted that a further improvement seems possible as the amount of data is crucial for training and the network's architecture of the asphere is more complex due to more input channels.

\subsection{Non-perfect system}
In any real world application, no experiment is carried out under perfect system conditions. This motivated the idea of disturbing the perfect simulated forward pass and of generalizing the model to non-perfect systems. The network now needed to cope with data coming from a non-perfect TWI after having trained on a perfect simulation environment in the first stage. This was achieved by using a conventional calibration to determine the correct model of the interferometer.

Here, we focused on the multi-spherical freeform artefact as the design topography. Thirty difference topographies were randomly chosen from the former test set, i.e. not included in U-Net training. They had a total root mean squared deviation of $545$ nm and are ranged from $296$ nm to $6.1\ \mu$m in their absolute maximal deviation from peak to valley. The results are shown in Table \ref{calib}, where the same trained network was used for differently produced inputs. The root mean squared error of the network predictions was $30$ nm on the perfect TWI system. This increased to $538$ nm after having disturbed the TWI system. The trained network is incapable of predicting properly. However, the error can be reduced to $67$ nm by using a calibrated forward pass to produce the input data. Hence, our proposed hybrid method can also generalize to non-perfect systems.
\begin{table}[h]
\centering
\begin{tabular}{cccc}
& \multicolumn{1}{c}{\textbf{Perfect system}} & \multicolumn{1}{c}{\textbf{\begin{tabular}[c]{@{}c@{}}Disturbed system,\\ no calibration\end{tabular}}} & \multicolumn{1}{c}{\textbf{\begin{tabular}[c]{@{}c@{}}Disturbed system,\\ with calibration\end{tabular}}} \\
\textbf{RMSE} & 30\ nm & 538\ nm & 67\ nm\\
\textbf{Median} & 16\ nm & 298\ nm & 33\ nm
\end{tabular}\caption{Root mean squared and median of absolute errors for the predictions of the same U-Net using different inputs. The perfect TWI system which was also used to generate the training data is in the first column, the disturbed TWI system without calibration is in the second column, and the hybrid method is in the third column.}\label{calib}
\end{table}

\section{Conclusion}
The obtained results are promising and suggest that deep learning can be successfully applied in the context of computational optical form measurements. The presented results are based on simulated data only and they constitute a proof-of-principle study rather than a final method that is ready for application. An extensive comparison with conventional methods is the next step. Testing the approach on real measurements and accounting for fine-tuning (such as the calibration of the numerical model of the experiment) is reserved for future work as well. Nevertheless, these initial results are encouraging and once trained, a neural network solves the inverse problem orders of magnitudes much faster than the currently applied conventional methods. We conclude from our findings that computational optical form metrology can also greatly benefit from deep learning.


\section{Acknowledgements}
The authors would like to thank Manuel Stavridis for providing the SimOptDevice software tool and Ines Fortmeier and Michael Schulz for their helpful discussions about optical form measurements.

\bibliography{ref}{}
\bibliographystyle{abbrv}
\end{document}